\documentclass[12pt]{article}
\usepackage[dvips]{graphics}

\begin{document}

	\begin{center}
	\LARGE{The masses of the mesons and baryons.\\
		Part V. The neutrino branch particles}
	\bigskip

	\Large{E. L. Koschmieder and T. H. Koschmieder}
	\medskip

	\small{Center for Statistical Mechanics, The University of Texas at 
Austin\\Austin, TX 78712, USA\\e-mail: koschmieder@mail.utexas.edu}
	\smallskip

	\large{April  2001}
	\end{center}

	\bigskip
	\noindent
	\small{We have determined theoretically the rest mass of the muon neutrino at 50 milli-eV and the rest mass of the electron neutrino at 5 meV, as well as , to 1\% accuracy, the ratio of the masses of the stable elementary particles which decay by weak decays. We assume that the particles of the neutrino branch consist of a cubic, isotropic nuclear lattice, held together by the weak nuclear force. The eigenfrequencies of the lattice are calculated with Born's theory of cubic lattices. Only neutrinos are required to explain the so-called stable particles of the neutrino branch.}

	\normalsize

	\section{Introduction}
We have shown in [1] that it follows from the well-known masses and the 
well-known decays of the so-called stable elementary particles that the 
stable particles consist of a $\gamma$-branch and a neutrino branch, the 
$\gamma$-branch particles decaying directly or ultimately into photons, 
whereas the neutrino branch particles decay directly or ultimately into 
neutrinos and e$^\pm$. The masses of the $\gamma$-branch, the $\pi^0$, 
$\eta$, $\Lambda$, $\Sigma^0$, $\Xi^0$, $\Omega^-$, 
$\Lambda^+_c$, $\Sigma^0_c$, $\Xi^0_c$ and $\Omega^0_c$ particles, are 
integer multiples of the mass of the $\pi^0$ meson, within at most 3.3\%, 
the average deviation being 0.0073. The masses of the neutrino branch, the 
$\pi^\pm$, K$^\pm$, n, D$^\pm$ and D$^\pm_S$ particles, are integer 
multiples of the mass of the $\pi^\pm$ mesons times a factor 
$0.86\,\pm\,0.02$. In [2] we have explained the integer multiple rule of 
the $\gamma$-branch particles with different modes and superpositions of 
plane, standing electromagnetic waves in a cubic nuclear lattice. For the 
explanation of the particles of the neutrino branch we follow a similar 
path. We assume, as appears to be quite natural, that the $\pi^\pm$ mesons 
and the other particles of the $\nu$-branch consist of the same particles 
into which they decay, that means of neutrinos and electrons or positrons. 
Since the particles of the $\nu$-branch decay through weak decays, we 
assume, as appears likewise to be natural, that the weak nuclear force 
holds the particles of the $\nu$-branch together. Since the range of the 
weak interaction is only about a thousands of the diameter of the 
particles, the weak force can hold particles together only if the 
particles have a lattice structure, just as macroscopic crystals are held 
together by microscopic forces between atoms. We will, therefore, 
investigate the energy which is contained in the oscillations of a cubic 
lattice consisting of electron and muon neutrinos. We will explain the 
ratios of the masses of the neutrino branch particles, in particular the 
factor $0.86\,\pm\,0.02$ which appears in the ratio of the masses of the 
$\nu$-branch particles divided by the mass of the $\pi^\pm$ mesons.

	\section{The frequency spectra of the oscillations of diatomic lattices}
Since we will explain the particles of the $\nu$-branch with the 
oscillations of a cubic lattice consisting of muon and electron neutrinos 
it is necessary to outline the basic aspects of diatomic lattice 
oscillations. In diatomic lattices the lattice points have alternately the 
masses m and M. The classic example of a diatomic lattice is the salt 
crystal with the masses of the Na and Cl atoms in the lattice points. The 
theory of diatomic lattice oscillations was developed by Born and v.Karman 
[3], referred to as B\&K. They first discussed a diatomic chain. The 
equation of motions in the chain are according to Eq.(22) of B\&K
\begin{eqnarray} m\ddot{u}_{2n} & = & \alpha(u_{2n+1}+u_{2n-1}-2u_{2n}) , \nonumber\\
                                                     & &\\
		 M\ddot{u}_{2n+1} & = & \alpha(u_{2n+2}+u_{2n}-2u_{2n+1}) , \nonumber
	\end{eqnarray}

\noindent
where the $u_n$ are the displacements, n an integer number and $\alpha$ a 
constant characterizing the force between the particles. As with any 
spring the restoring forces in (1) increase with increasing distance 
between the particles. Eq.(1) is solved with
	\begin{eqnarray}u_{2n} & = & Ae^{i(\omega t+2n\phi)} ,\nonumber\\
			      &     & \\
		u_{2n+1} & = & Be^{i(\omega t+(2n+1)\phi)} ,\nonumber
	\end{eqnarray}

	\noindent	   
where A and B are constants and $\phi$ is given by
	\begin{equation}\phi = 2\pi a/\lambda  ,\end{equation}	
\noindent
where $a$ is the lattice constant and $\lambda$ the wavelength, $\lambda= 
na$. The solutions of Eq.(2) are obviously periodic in time and space and 
describe standing waves in which for all times t the displacements are 
zero at the nodes 2n$\phi$ = $\pi$/2 and (2n+1)$\phi$ = $\pi$/2. Using (2) 
to solve (1) leads to a secular equation from which according to Eq.(24) 
of B\&K the frequencies of the oscillations of the chain follow from 

	\begin{equation} 4\pi^2\nu^2 = \frac{\alpha}{Mm}\cdot(M+m \pm 
\sqrt{(M-m)^2+4mMcos^2\phi}) . 
	\end{equation} 
\noindent
Longitudinal and transverse waves are distinguished by the minus or plus 
sign in front of the square root in (4). We will encounter later on a 
similar equation for the plane waves in an isotropic three-dimensional 
lattice. The equations of motion for the oscillations in a 
three-dimensional diatomic lattice have been developed by Thirring [4].

	\section{The neutrino lattice oscillations}
The particles of the neutrino branch decay primarily by weak decays, see 
Table 2 in [1]. The characteristic case are the $\pi^\pm$ mesons which 
decay via e.g. $\pi^+\rightarrow\mu^++\nu_\mu$ (99.988\%) followed by  
$\mu^+\rightarrow e^+ +\bar{\nu}_\mu+\nu_e$ ($\approx$100\%). Only neutrinos 
result from 
the decay of the $\pi^\pm$ mesons, but for $e^\pm$ which conserve charge. 
If the particles consist of the particles into which they decay, then the 
$\pi^\pm$ mesons and the other particles of the neutrino branch are made 
of neutrinos and $e^\pm$. The neutrinos must be held together in some 
form, 
otherwise the particles could not exist over a finite lifetime, say 
$10^{-10}$ sec. Since neutrinos interact through the weak force which has 
a range of order of $10^{-16}$ cm according to p.25 of [5], and since the size of 
the nucleon is of order of 10$^{-13}$ cm, the only apparent way for the 
weak force to hold the particles together is through a neutrino 
lattice.                   We suggest that the lattice is cubic, because a 
cubic lattice is held together by central forces, which are the most 
simple forces to consider. It is not known with certainty that neutrinos 
actually have a rest mass as was originally suggested by Bethe [6] and 
Bahcall [7] and what the values 
of m($\nu_e$) and m($\nu_\mu$) are. However, the results of the 
Super-Kamiokande experiments [8] indicate that the neutrinos have rest 
masses and that the difference of both rest masses is about 70 milli-eV. 
The neutrino lattice must be diatomic, meaning that the lattice points 
have alternately larger (m($\nu_\mu$)) and smaller (m($\nu_e$)) masses. We 
will retain the traditional term diatomic. The lattice we consider is 
shown in Fig. 1.

	\begin{figure}[h]
	\vspace{0.5cm} 
	\hspace{2cm}
	\includegraphics{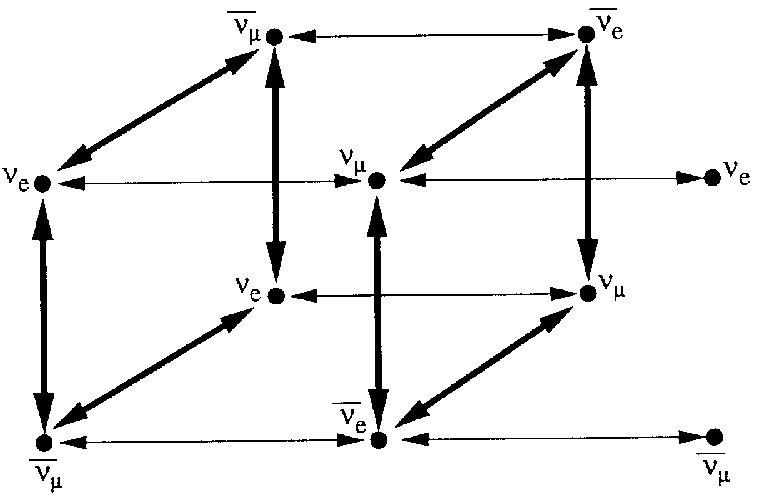}
	\vspace{-0.2cm}
	\begin{quote} 
Fig.1: The neutrino lattice. Bold lines mark the forces between neutrinos and antineutrinos. Thin lines mark the forces between either neutrinos only, or antineutrinos only.
	\end{quote}
	\end{figure}

Since the neutrinos have spin $\frac{1}{2}$ this is a four-Fermion 
lattice as is required for Ferm\i's explanation of the $\beta$-decay. The 
first investigation of cubic Fermion lattices in context with the 
elementary particles was made by Wilson [9]. The entire neutrino lattice 
is electrically neutral and has no spin, the spin of each neutrino is 
canceled by the opposite spin of an adjacent neutrino. Since we do not 
know the structure of the electron we cannot consider charge.

   A neutrino lattice takes care of the continuum of frequencies which 
must, according to Fourier analysis, be present after the high energy 
collision which created the particle. The continuum of frequencies of the 
Fourier spectrum can be absorbed by the continuous frequency spectrum of 
the lattice oscillations. We will, for the sake of simplicity, first set 
the sidelength of the lattice at $10^{-13}$ cm that means approximately 
equal to the size of the nucleon. The lattice then contains $10^9$ 
lattice points, since the lattice constant $a$ is of order $10^{-16}$ cm. 
The 
sidelength of the lattice does not enter Eq.(6) for the frequencies of the 
lattice oscillations. The calculation of the ratios of  
the energy of different lattice oscillations is consequently independent of the 
size of the lattice. However, the size of the lattice can be determined 
theoretically as will be shown later.  

   We require that the lattice is isotropic which is the most simple 
possible case. Isotropy means that the ratio of the force constant 
$\alpha$ for the lattice points at distance $a$ to the force constant 
$\gamma$ for the lattice points at distance $a\sqrt{2}$ is 
$\gamma/\alpha = 0.5$.  That $\gamma/\alpha$ = 0.5, in the isotropic 
case, follows from the definition of $\gamma$ and $\alpha$.  According to 
Eq.(13) of B\&K it is 
	\begin{eqnarray} \alpha & = & a(c_{11}-c_{12}-c_{44}) ,\nonumber\\
4\gamma & = & a(c_{44}+c_{12} ),
	\end{eqnarray}   
	\noindent                   
where c$_{11}$, c$_{12}$ and c$_{44}$ are the elastic constants of an 
elastic
body in continuum mechanics which applies in the limit $a\rightarrow 0$.  
The neutrino lattice is actually the best approximation to an elastic body 
of continuum mechanics. If we consider central forces then $c_{12} = c_{44}$ 
which is the classical Cauchy relation. Isotropy requires that $c_{44} = (c_{11} - c_{12})/2$ or that in this case $3c_{44} = c_{11}$.

   Using Thirring's equations of motion for a diatomic three-dimensional 
lattice and assuming that the displacements u$_{l,m}$ of the smaller 
masses and U$_{l+1,m}$ of the larger masses are periodic in space and 
time, similar to Eq.(2), one arrives at a characteristic equation from 
which after a very long but straightforward calculation the frequency 
equation for the two-dimensional, i.e. plane oscillations of a cubic, 
isotropic, diatomic lattice is obtained. It is

	\begin{eqnarray}
 \nu^2_\pm = \frac{\alpha}{4\pi^2 M} 
  \cdot \left[ \hspace{.2cm} \left(\frac{M}{m}+1\right)  \left(1+\frac{2\gamma}{\alpha} 
(1-\cos\phi_1\cos\phi_2 + \sin\phi_1\sin\phi_2) \right) \right.\nonumber\\*[.5cm]
 \left. \pm\, \sqrt{\left( \frac{M}{m}-1\right)^2  
\left(1+\frac{2\gamma}{\alpha} \, 
(1-\cos\phi_1\cos\phi_2+\sin\phi_1\sin\phi_2)\right)^2 +\frac{4M}{m}   
\cos^2\phi_1} \hspace{.3cm} \right] \nonumber\\*
	\end{eqnarray} 

\noindent	
This is a two-dimensional version of Eq.(4) above which applied to a 
diatomic chain. In Eq.(6) are, for each pair $\phi_1,\phi_2$, two 
different frequencies $\nu_-$ or $\nu_+$ caused by the $\pm$ sign in front 
of the square root. These frequencies correspond to longitudinal (minus 
sign) and transverse oscillations (plus sign). It is important to note 
that each value of $\nu_+$ and $\nu_-$ comes with a plus as well as with a 
minus sign, because Eq.(6) determines the square of $\nu_+$ and $\nu_-$.

   From Eq.(6) we can calculate the frequency distribution for all values 
-$\pi\leq \phi_1,\phi_2\leq\pi$. These values depend on the ratio M/m, 
where M is the mass of the muon neutrino and m the mass of the electron 
neutrino. The distribution of $\nu_-/\nu_0$ following from Eq.(6) with $\nu_0 = \sqrt{\alpha/4\pi^2M}$ is shown in Fig.\ 2 for the range $0\leq\phi_1,\phi_2\leq\pi$.
  
	\begin{figure}[h]
	\vspace{0.5cm}
	\hspace{3cm} 
	\includegraphics{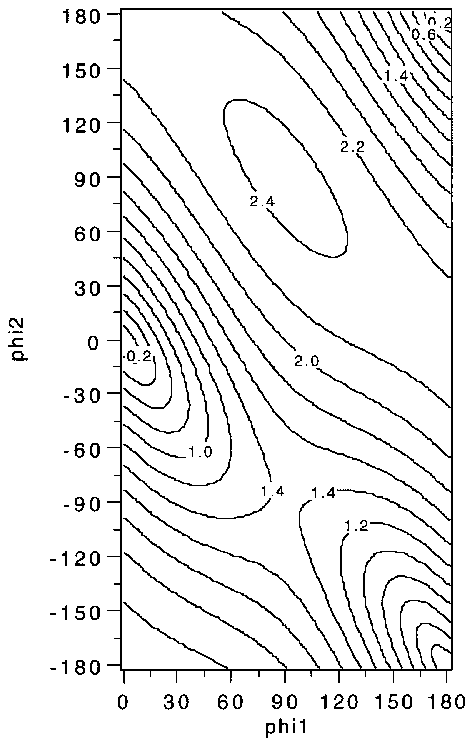}
	\vspace{-0.3cm}
	\begin{quote}
Fig.2: The frequency distribution $\nu_-/\nu_0$ of the basic mode 
according to Eq.(6). Diatomic, isotropic case with M/m = 9.3.
	\end{quote}
	\end{figure}

\noindent
The value of M/m 
originates from a trial run. There are some easily verifiable values of  the frequencies
$\nu_-$  of the longitudinal waves.
 At $\phi_1,\phi_2 = 0,0$ or $\pi,\pi$, or $\pi,-\pi$  it must be that 
$\nu_- = 0$. At $\phi_1,\phi_2 = \pi/2,\pi/2$ it must be that $\nu_- = 
\nu_0\sqrt{6}$. At $\phi_1,\phi_2 = \pi/2,-\pi/2$ it must be that $\nu_- = 
\nu_0\sqrt{2}$.

   From the frequency distribution follows the group velocity $d\omega/dk$ 
at 
each point $\phi_1,\phi_2$. With $k = 2\pi/\lambda$ and $\phi = 2\pi 
a/\lambda$ it follows that 

\begin{equation} d\omega/dk = 2\pi a d\nu/d\phi .\end{equation}
\noindent
With $\nu = \nu_0 f(\phi_1,\phi_2)$ we have

\begin{equation} c_g = d\omega/dk = a\sqrt{\alpha/M}\cdot 
df(\phi_1,\phi_2)/d\phi .\end{equation}

In order to determine the value of $d\omega/dk$ we have to know the value 
of $\sqrt{\alpha/M}$. From Eq.(5) for $\alpha$ follows that $\alpha = 
ac_{44}$ in the isotropic case with central forces, when $c_{12} = c_{44}$ 
(Cauchy's relation) and $c_{11} = 3c_{44}$. The group velocity is therefore

\begin{equation} c_g = \sqrt{a^3c_{44}/M} \cdot df/d\phi = 
\sqrt{c_{11}/{3\rho}} \cdot df/d\phi .\end{equation}
\noindent
We now set $a\sqrt{\alpha/M} = c_*$, where $c_*$ is the velocity of 
light, which is a 
necessity because the neutrinos in the lattice soon approach the velocity 
of light. In a formula 

\begin{equation} c_* = \sqrt{a^3c_{11} /3M} = a\sqrt{\alpha/M} 
.\end{equation}
\noindent
It follows that 
\begin{equation} c_g = c_* \cdot df/d\phi ,\end{equation}  
\noindent
as it was with the $\gamma$-branch in [2]. Since $c_g$ must be $\leq c_*$  
Eq.(11) limits 
the value of $df(\phi_1,\phi_2)/d\phi$ to $\leq 1$, regardless whether we 
consider $\nu_-$ or $\nu_+$. There is no extra spectrum for transverse 
oscillations.
 
    From Eq.(10) follows the mass of the muon neutrino, it is
 
\begin{equation} M = a^3c_{11}/3c_*^2 .\end{equation}
\noindent 
According to Eq.(18) of [10] $c_{11}$ can be determined theoretically from 
an 
exact copy of the determination of $c_{11}$ in Born's lattice theory, 
assuming 
weak nuclear forces between lattice points at distance $a = 10^{-16}$ cm, 
and 
replacing $e^2$ by $g^2$, where $g^2$ is the interaction constant of the 
weak 
force. We note that the value of $c_{11}$ in [10] is, within the 
uncertainty of 
its determination, compatible with a completely different theoretical 
determination of the compression modulus of the nucleon [11]. Using
the equation for the elasticity constant $c_{11} = 0.538g^2 
\epsilon/a^4$ from Eq.(17) 
of [10], and with $ \epsilon = 1.5\cdot10^{-12}$ from Eq.(10) of 
[10], we find that

\begin{equation} M = m(\nu_\mu) = 0.538g^2 \epsilon/3ac_*^2 = 
2.69\cdot10^{-13}g^2/ac_*^2 , \end{equation}
\noindent
which shows that $m(\nu_\mu)c_*^2$  depends only on the weak interaction 
constant  
$g^2$ and the range $a$ of the weak force. Using now the same values for $a 
= 
10^{-16}$ cm  and $g^2 = 2.946\cdot10^{-17}$ erg cm we used before 
in [10] we find 

\begin{equation} M = m(\nu_\mu) = 0.9\cdot10^{-34} gr = 5\cdot10^{-2} 
eV/c_*^2 .\end{equation}

   Within the accuracy of the parameters $g^2$ and $a$ the mass of the muon 
neutrino is 50 milli-electron-Volt. It can be verified easily that the 
mass of the muon neutrino we have found makes sense. The energy of the 
rest mass of the $\pi^\pm$ mesons is 139 MeV, and we have about $10^9$ 
lattice 
points. Therefore there can be on the average no more energy per lattice 
point than about 0.14 eV. According to Eq.(14) about 50\% of the available 
energy per lattice point goes into a muon neutrino mass, a small part, as 
we will see, goes into an electron neutrino mass, the rest goes into the 
lattice oscillations.

	\section{The consequences of the rest masses of the neutrinos}
If the neutrinos have a rest mass and a continuous frequency distribution then the masses of the particles of the neutrino branch 
follow from the formula

\begin{equation} m(n) = \sum_{i=1}^N [ \frac{m(\nu_\mu)_0}{\sqrt{1-\beta^2_{i\mu 
n}} }+ \frac{m(\nu_e)_0}{\sqrt{1-\beta^2_{ien}}}] . \end{equation} 

The $\beta_i = v_i/c_*$ are the time averages of the velocity of each 
oscillation, 
the index n goes from 1 to 5, n = 1 marking the $\pi^\pm$ mesons and n = 5 
the 
$D_S^\pm$ mesons. The $\beta_{i\mu n}$ of the muon neutrinos are different 
from the                                  $\beta_{ien}$ of the electron 
neutrinos, both neutrino types oscillate with the same 
frequencies which means that they have the same energies $E_i = h\nu_i$ 
but 
since $m(\nu_e) <  m(\nu_\mu)$ the $\beta_i$ of each electron neutrino is 
larger than the $\beta_i$ 
of the corresponding muon neutrino. The different velocities reflect the different displacements of both neutrino types. In order to evaluate the sums in 
Eq.(15) we have to know the number of muon or electron neutrinos in the 
lattice. Recent measurements and theoretical analysis put the value of the 
proton radius at $r_p = (0.88 \pm  0.015)\cdot10^{-13}$ cm [12,13]. The 
radius of the $\pi$ 
mesons from experiments is given as $(0.74 \pm 0.03)\cdot10^{-13}$ cm 
[14]. 
Theoretical analysis raises the pion radius to $r_\pi = 0.83\cdot10^{-13}$ 
cm [15]. In 
the following we use $r_p = r_\pi = 0.88\cdot10^{-13}$ cm. From this 
follows with $a = 
10^{-16}$ cm that the number of the muon as well as of the electron 
neutrinos 
in the cubic nuclear lattice is $N = 1.427\cdot10^9$ or that the total 
number of 
neutrinos in the lattice is $2N = 2.854\cdot10^9$.

    Since we cannot determine the velocity of each of the $10^9$ particles of 
the lattice we determine the maximal value of the velocity of the 
neutrinos in the lattice using the well-known fact that the sum of the 
potential energy and the kinetic energy of a harmonic oscillator is 
constant and equal to the maximal kinetic energy. Averaged over i and 
time, we have in the case of the muon neutrinos, for the particle with 
index n

\begin{equation} Nm(\nu_\mu)_0\cdot[1/\sqrt{1-\beta_n^2}-1] =
			1/2\cdot (m(n) - N[m(\nu_\mu)_0 + m(\nu_e)_0]) ,\end{equation}
\noindent
with a corresponding equation for the electron neutrinos. The sum of the 
maximal kinetic energies or the sum of potential and kinetic energy of the 
muon and electron neutrinos is, according to Eq.(16), equal to the 
difference between the rest mass of the particle m(n) and the sum of the 
rest masses of both neutrino types. The $\beta$ of the particles of the 
neutrino 
branch with N = 0.6$\cdot10^9$ are listed in Table 1. For $m(\nu_\mu)_0$ we have used 50 meV 
according to 
Eq.(14), and for $m(\nu_e)_0$ we have used 5 meV, which value will be 
justified 
soon from the ratios of the masses m(n)/m($\pi^\pm)$. Replacing the many 
$\sqrt{1-\beta_{in}^2}$ in Eq.(15) with the averages $\sqrt{1-\beta_n^2}$ from 
Eq.(16) we obtain correctly the masses of the different particles m(n).

	\begin{table}[h]
		\begin{tabular}{llllll} \hline \hline 
particle &     $\pi^\pm$ &  $K^\pm$ &       n&     $D^\pm$&      $D_S^\pm$\\ [.5ex]
		\hline
 index n &       1&         2&        3&          4&          5\\
 mode ($i_1,i_2$)&  (1.1)&     (2.2)&     2$\cdot$(2.2)&   4(2.2)&  2(2.2) + (3.3)\\ 
$\beta_\mu$&            0.9329&    0.9933&    0.9981&    0.9995&     0.99955\\
 $\beta_e$&              0.9986&     0.99992&    0.99998&    0.99999&     0.99999\\
 E(n)/E(1)&     1&         0.8713\hspace{.1cm}$\cdot$ 4&  0.8467\hspace{.1cm}$\cdot$ 8&  0.8328\hspace{.1cm}$\cdot$ 16&  0.8321\hspace{.1cm}$\cdot$ 17\\
 m(n)/m($\pi^\pm$) &    1&         0.8843\hspace{.1cm}$\cdot$ 4&  0.8415\hspace{.1cm}$\cdot$ 8&  0.8371\hspace{.1cm}$\cdot$ 16&  0.8296\hspace{.1cm}$\cdot$ 17\\
			\hline \hline
		\end{tabular}
	\caption{The average velocities and the calculated and              
experimental mass ratios of the neutrino branch particles.}
	\end{table}

   We will now determine the mass ratios of the $\nu$-branch particles 
from 
the frequencies of the neutrino lattice oscillations. The actually 
possible oscillations are given by Eq.(11), in which $df/d\phi$ must be 
$<$ 1, 
because the neutrinos cannot move with the velocity of light, $c_g$ is 
then 
$\beta c$. The possible frequency distributions are like Fig. 3 only that 
the 
slope is $df/d\phi = \beta$, not 1  as in the $\gamma$-branch. The 
frequency distributions 
of the higher modes are very similar to Fig. 3  because $\beta$ of the 
higher 
modes is soon practically equal to one, as shown on Table 1. 
	\begin{figure}[t]
	\unitlength1cm
		\begin{minipage}[t]{6.5cm}
		\includegraphics{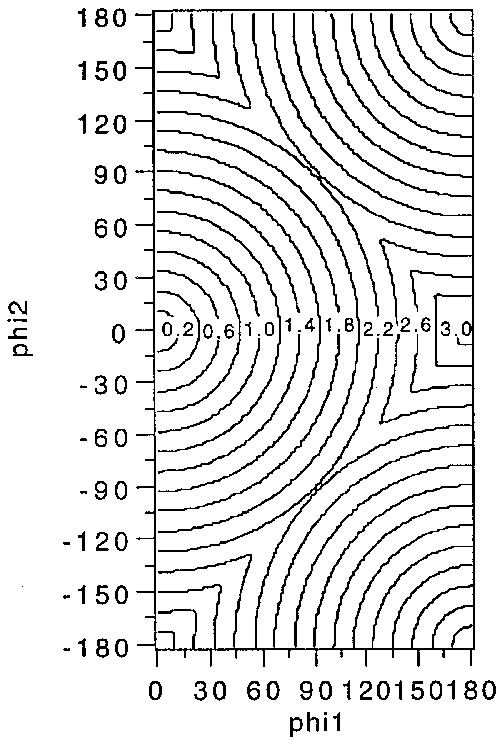}
		\begin{quote}
Fig.3: The frequency distribution $\nu/\nu_0$ of the basic (1.1) mode,
according to Eq.(11), with slope 1.
		\end{quote}
		\end{minipage}
			\hfill
		\begin{minipage}[t]{6.5cm}
		\includegraphics{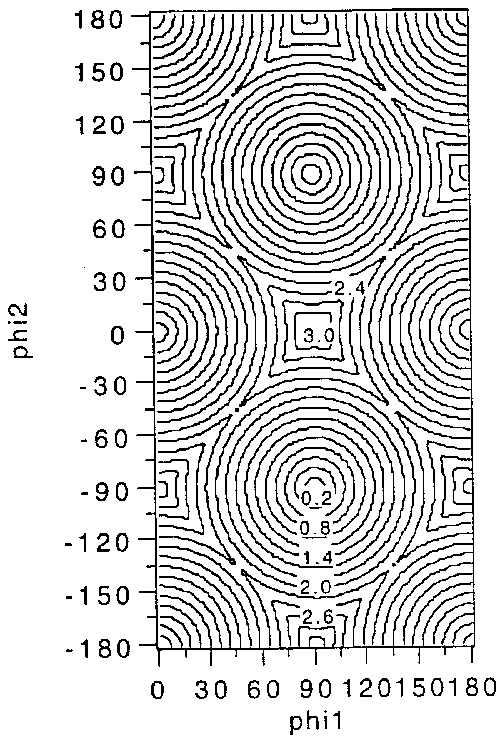}
		\begin{quote}
Fig.4: The frequency distribution $\nu/\nu_0$ of the (2.2) mode, according 
to Eq.(11), with slope 1. The variables $\phi_1$ and $\phi_2$ are 
one-half of the actual $\phi$ of the second mode.
		\end{quote}
		\end{minipage}
	\end{figure}
\noindent
The higher 
modes of the oscillations are obtained by replacing the variables $\phi_1$ 
and 
$\phi_2$ in the ansatz for the solutions of the diatomic lattice 
oscillations by 
the variables $i_1 \phi_1$ and $i_2 \phi_2$, where $i_1$ and $i_2$ are 
integers. 
In all cases considered $i_1 = i_2$. For $\phi> 0$ the function 
$f(\phi_1,\phi_2)$ extends then, for 
example, from 0 to 2$\pi$ for the first higher mode ($i_1,i_2 = 2)$. The 
frequency 
distribution of the first higher mode with slope 1 is shown in Fig.\ 4. The 
area covered by Fig.\ 4 is four times the area covered by the first mode 
(Fig. 3), consequently the number of oscillations is four times the number 
of oscillations of the first mode, but the frequencies are the same as 
those of the first mode.

   The energy contained in all longitudinal lattice oscillations is given 
by
  	
\begin{equation} E =
\frac{Nh\nu_0}{(2\pi)^2} \int\int\limits_{-\pi}^{\pi} \nu_-/\nu_0 (\phi_1,\phi_2) 
d\phi_1d\phi_2 .\end{equation}
\noindent 
This equation was used before in [2], it originates from Eq.(50) of B\&K. 
The numerical value of the double integral in Eq.(17) of the (1.1) state 
shown in Fig.\ 3 was found by numerical integration to be 66.9 
$\mathrm{radians}^2$. If 
the frequencies of the oscillations in the neutrino lattice increase with 
$\beta$ the value of the double integral in Eq.(17) has to be multiplied by 
$\beta$ too. 
The ratio of the energy of the oscillations of a state ($i_n,i_n$) to the 
basic state (1.1) is then given by $i_n^2 \beta_n/\beta_1$ or by the sum  
of different $i_n^2$  times $\beta_n/\beta_1$ if a particle is the 
superposition of 
different modes. The factor $i_n^2$ originates from the increased area of 
the 
higher modes, see e.g. Fig.\ 4. 
   The particles of the $\nu$-branch are, however, not made up exclusively 
by 
the energy of the neutrino lattice oscillations but also by the energy 
contained in the neutrino rest masses. From $E(n) = m(n)c_*^2$ = $E_k + 
N[m(\nu_\mu)_0 + m(\nu_e)_0]c_*^2$, where $E_k$ is the kinetic or oscillation energy, 
follows that

\begin{equation}  \frac{E(n)}{E(1)} = \frac{E_k (n)}{E_k (1)} 
\cdot\frac{1+\Sigma/E_k(n)}{1+\Sigma/E_k(1)} , \end{equation}

\noindent													
where we have abbreviated the sum over the neutrino rest masses by 
$\Sigma$.  In 
the case of the  photons of the $\gamma$-branch particles  $\Sigma$ is 
zero and 
$\beta_n$ = 1, so $E_k(n)/E_k(1)$ is simply $i_n^2$ or equal to the sum of 
the $i_n^2$ of 
different modes, that means the integer multiple rule of the 
$\gamma$-branch applies.

 In the case of the neutrino branch, where the 
neutrinos have 
rest masses, the ratio $E(n)/E(1)$ is not an integer number and is not 
constant, but is determined by $i_n^2$ times a rest mass factor $R$ 

\begin{equation} \frac{E(n)}{E(1)} = i_n^2 R = 
i_n^2\cdot\frac{\beta_n}{\beta_1} 
(\frac{1+\Sigma/E_k(n)}{1+\Sigma/E_k(1)} ) \end{equation}

\noindent
or, if different modes are involved, $i_n^2$ is replaced by the sum of 
different $i_n^2$. In order to determine $\Sigma/E_k(n)$ we use the 
empirical values 
of $E_k(n)$ which are the energies of the particles $m(n)$ minus the sum of 
the 
energies of the rest masses of the neutrinos. Since $E_k(1)$ is smaller 
than 
$E_k(n>1)$ and since $\Sigma$ = const it follows that the term in 
parenthesis in 
$R$ is $<$ 1 and decreases with increased n, whereas $\beta_n/\beta_1$ is $ >$ 
1 and 
increases as $\beta_n$ approaches 1. That means that the ratio $E(n)/E(1)$ 
is an 
integer number times the rest mass factor which depends on n and decreases 
as n increases. This is so because the contribution of the rest masses of 
the neutrinos to the energy of the particle decreases. In the asymptotic 
case of large n the rest mass factor approaches 0.82. The different values 
of these factors are listed in Table 1 on the next to last line and can be 
compared to the factors following from the measured masses of the 
$\nu$-branch 
in the last line of Table 1. A complete agreement of the calculated and 
experimental factors cannot be expected because we do not take into 
account the consequences of spin, isospin, strangeness and charm. The 
existence of the rest mass factors in the ratios of the $\nu$-branch particle masses is 
solely a consequence of the rest masses of the neutrinos.

   The ratio of the masses of the $\nu$-branch is a weak function of the 
mass 
of the electron neutrino which enters Eq.(19) through $\Sigma$  of which 
$\Sigma m(\nu_e)$ 
is the smaller part. The value of $m(\nu_e)$ = 5meV used in the 
calculations 
leading to the numbers in Table 1 has been determined by trial and error, 
it is the $m(\nu_e)$ which produced the smallest deviation ($\pm$ 0.7\%) of 
the 
calculated rest mass factors to the experimentally found factors on the 
last line of Table 1. Only integer values of meV have been considered 
because the uncertainty of $m(\nu_\mu)$ does not warrant the consideration 
of 
fractions of meV.

   As in the case of the particles of the $\gamma$-branch we have also 
found the 
antiparticles of the $\nu$-branch. From Eq.(6) follows that for each 
positive 
frequency there exists also a negative frequency with the same absolute 
value. That means, as follows from Eq.(17), that the energy contained in 
all lattice oscillations is negative if the frequencies are negative. 
Since the total energy of a $\nu$-branch particle consists of the energy 
of 
the lattice oscillations plus the energy of the rest masses of the 
neutrinos, we have to show that the energy of the neutrino rest masses can 
be negative, in other words that there are antineutrinos with the same 
absolute mass as those of the neutrinos. Only then will the absolute value 
of the mass of an antiparticle of the $\nu$-branch be equal to the mass 
of the 
corresponding particle, as must be.

   The existence of the muon antineutrinos follows from Eq.(7) for the 
group velocity of the waves in the neutrino lattice. Since $c_g = 2\pi a d\nu 
/d\phi$  and since 
the slope of the negative frequencies $d\nu_{an}/d\phi$ in the antiparticle 
has the 
opposite sign of the slope $d\nu/d\phi$ in the particle and since $c_g$ is 
the same 
in both cases it must be that the lattice distance $a_{an}$ of the 
antiparticles has the opposite sign of the lattice distance $a$ of the 
particles, but has the same absolute value as $a$. That means that the 
coordinate system of the antineutrinos is turned by $180^0$ around a 
central
axis perpendicular to the plane waves in the neutrino lattice. If $a_{an} = 
-a$ 
then the mass of a muon antineutrino is negative according to Eq.(13), 
but has the same absolute value as the muon neutrino. Since we do not have 
a formula for the mass of the electron neutrino we cannot show that 
electron antineutrinos exist as well. But, since the antiparticles of the 
$\nu$-branch have the same mass as the $\nu$-branch particles, 
electron antineutrinos with the same absolute mass as the electron 
neutrinos must exist.

   The size of the particles does not enter Eq.(6) for the frequencies of 
the lattice oscillations, so the particle size has to be determined 
separately. As we have shown in [16] the particle size of the 
$\gamma$-branch 
particles is determined by the radiation pressure. The particle will break 
up at the latest if the outward directed radiation pressure is equal to 
the inward directed force between one lattice layer and the adjacent 
layer. A similar consideration applies for the $\nu$-branch particles. The 
pressure in the neutrino lattice $p = \sum_{i=1}^N 
[m(\nu_\mu)\beta_\mu^2c_*^2 + m(\nu_e)\beta_e^2c_*^2]$ is 
proportional to the size of the lattice because N is proportional to the 
lattice size. If the outward directed pressure is equal to the force between the 
lattice layers, in technical terms to the Youngs modulus of the lattice, 
the lattice will break up, at the latest. Details cannot be discussed here.

   Now we can explain the particles of the neutrino branch. The basic 
state $i_1,i_2$ = (1.1) of the neutrino lattice oscillations corresponds 
to 
the $\pi^\pm$ mesons. The first higher mode $i_1,i_2$ = (2.2) corresponds 
to the $K^\pm$ 
mesons. A superposition of two coupled (2.2) modes produces the neutron 
which has spin $\frac{1}{2}$, just as the superposition of two coupled $\eta$ mesons 
or 
(2.2) modes produces the $\Lambda$ particle with spin $\frac{1}{2}$ in the 
$\gamma$-branch. The 
superposition of a proton and a neutron or of two neutrons with opposite 
spin creates the 
$D^{\pm,0}$ meson. The superposition of two (2.2) states on a (3.3) state 
creates 
the $D_S^\pm$ meson. All $\nu$-branch particles are accounted for, their 
masses are 
integer multiples of the basic state times the rest mass factor which is 
of order 0.85. The agreement of the calculated masses with the 
experimentally determined masses is in the 1\% range.

	\section{Conclusions}

We have shown that the masses of the neutrino branch of the so-called 
stable elementary particles, the $\pi^\pm, K^\pm, n, D^\pm$, and $D_S^\pm$ 
particles, can be 
explained as the sum of the energies of the oscillations of plane, standing waves 
in a cubic, diatomic, isotropic neutrino lattice and the rest masses of 
the neutrinos. In particular, we have explained, as a consequence of the 
rest masses of the neutrinos, the factors of about 0.85 which appear in 
the ratios of the $\nu$-branch particles to the $\pi^\pm$ mesons. We have 
also 
determined the rest masses of the muon neutrino and the electron neutrino. 
Both neutrino types as well as the particles of the $\nu$-branch have 
automatically antiparticles.

   Let us consider what we have learned from the standing wave model we 
propose. We started with the elementary observation that the so-called 
stable mesons and baryons consist of a $\gamma$-branch and a neutrino 
branch, 
and that the masses of the $\gamma$-branch are, in a good 
approximation, 
integer multiples of the $\pi^0$ meson. We have also observed that, 
according to 
Fourier analysis, a continuum of frequencies must be present in a particle 
created by a high energy collision. This continuum of frequencies can be 
absorbed by the oscillations of a nuclear lattice. We have explained the masses of the particles and antiparticles of the $\gamma$-branch with different modes and superpositions of modes of the frequency distributions of a nuclear lattice, using nothing else but photons. The masses so determined 
agree on the average within 0.7\% with the mass ratios of the measured 
particles masses, an accuracy not matched in the theory of the particles. 
The remaining stable particles, the particles of the $\nu$-branch, can also be explained 
by 
the standing wave model as we have shown above. In this case we consider 
the oscillations of a lattice consisting of muon and electron neutrinos. 
We have determined the rest masses of the muon and electron neutrinos, as 
well as the ratios of the masses of the particles 
 of the $\nu$-branch. In other words we have, except for the 
bottom particles, determined the masses of the entire spectrum of stable 
mesons and baryons using only photons or neutrinos.

 We note that a nuclear lattice automatically entails the existence of a strong attractive force between the sides of two lattices, resulting from the unsaturated forces of the $10^6$ lattice points at each side. This force can be calculated following Born and Stern [17] as discussed in [10]. We will show in a 
forthcoming paper that the spin of the particles can also be explained 
with standing waves in the particles.

\bigskip
    We are grateful to Professor I. Prigogine for his support.   
         
\bigskip

\noindent
\textbf{REFERENCES}
\smallskip

\noindent
[1] Koschmieder,E.L. Bull. Acad. Roy. Belgique, {\bfseries X},281 (1999).\\
\indent
    arXiv: hep-ph/0002179 (2000), arXiv: hep-ph/0011158 (2000).    
\smallskip

\noindent
[2] Koschmieder,E.L. and Koschmieder,T.H.\\
\indent  Bull. Acad. Roy.   Belgique, {\bfseries X},289 (1999).\\
\indent
 arXiv: hep-lat/0002016 (2000).
\smallskip

\noindent
[3] Born,M. and v.Karman,Th. Phys. Z. {\bfseries 13},297 (1912).
\smallskip

\noindent
[4] Thirring,H. Phys. Z. {\bfseries 15},127 (1914).
\smallskip

\noindent
[5] Perkins,D.H. \textit{Introduction to High Energy Physics},\\
\indent
 Addison-Wesley (1982).
\smallskip

\noindent
[6] Bethe,H. Phys. Rev. Lett. {\bfseries 58},2722 (1986).
\smallskip

\noindent
[7] Bahcall,J.N. Rev. Mod. Phys. {\bfseries 59},505 (1987).
\smallskip

\noindent
[8] Fukuda,Y. et al. Phys. Rev. Lett. {\bfseries 81},1562 (1998).
\smallskip

\noindent
[9] Wilson,K.G. Phys. Rev. D {\bfseries 10},2445 (1974).
\smallskip

\noindent
[10] Koschmieder,E.L. Nuovo Cim. {\bfseries 101},1017 (1989).
\smallskip

\noindent
[11] Badhuri,R.K., Dey,J. and Preston,M.A. \\
\indent
Phys. Lett. B. {\bfseries 136},289 (1984).
\smallskip

\noindent
[12] Rosenfelder,R. arXiv: nucl-th/9912031 (2000).
\smallskip

\noindent
[13] Karshenboim,S.G. Can. J. Phys. {\bfseries 77},241 (1999).
\smallskip

\noindent
[14] Liesenfeld,A. et al. Phys. Lett. B {\bfseries 468},20 (1999).
\smallskip

\noindent
[15] Bernard,V., Kaiser,N. and Meissner,U-G. \\
\indent
arXiv: nucl-th/0003062 (2000).
\smallskip

\noindent
[16] Koschmieder,E.L. arXiv: hep-lat/0005027 (2000).
\smallskip

\noindent
[17] Born,M. and Stern,O. Sitzungsber. Preuss. Akad. Wiss. {\bfseries 33},90 (1919).

\end{document}